# MONITORING THE ENERGY CONSUMED BY A NETWORK INFRASTRUCTURE TO DETECT AND ISOLATE FAULTS IN COMMUNICATION ARCHITECTURE

## Dimitar Minovski[1], Eric Rondeau[2,3] and Jean-Philippe Georges[2,3]

[1] Université de Lorraine, Vandœuvre-lès-Nancy, France
[2] Université de Lorraine, CRAN, UMR 7039, Campus Sciences, BP 70239, Vandœuvre-lès-Nancy Cedex, 54506, France
[3] CNRS, CRAN, UMR 7039, France

## ABSTRACT

*In recent years, a number of major improvements were introduced in the area of computer networks, energy-efficient network protocols and network management systems. Software Defined Networking (SDN) as a new paradigm for managing complex networks brings a significant opportunity to reduce the energy consumption among ICT. In this paper, we are tackling improvements in the process of monitoring the states of the networking devices and optimizing the existing solutions. We are monitoring the energy consumption of a network architecture and augment the retrieved raw power data to detect changes in the state of the devices. The goal is to benchmark the difference between the power data fetched from the real-measures and the data extracted from the power models, translated as the expected behavior of the devices. An application is designed to monitor and analyze the retrieved power data of a simulated ICT infrastructure composed of Cisco switches and routers, Dell Precision stations and Raritan PDUs. Moreover, smart algorithms are developed which outcome results with detection of changes in the state of the devices, as well as with detection and isolation of possible anomalies and their impact on the energy consumption. The application is envisioned to be an extension to the existing SDN controllers for monitoring and reporting the changes in the state of the ICT devices.*

# INTRODUCTION

The energy-efficient network infrastructures have recently become a hot topic in the business world, as the concept of Green IT strives to reduce the overall operational costs, but in the same time also to eliminate the inefficiencies from the enterprises' IT systems. The network infrastructures are already massively deployed and even projected to have exponential growth (Cisco, 2011) due to the evolution of the Internet, user demands and trending topics such as Internet of Things (Ovidiu and Friess, 2013). Thus, as a shared resource they have to be constantly available, which exacerbates the sustainability issues. Researchers have already proposed different network-wide energy management schemes targeting various areas such as datacenters (Koutitas, 2010), mobile networks (Wang et al., 2012) and WANs (Gupta and Singh, 2003). However, it is quite challenging to tackle the energy efficiency issues within the households and from small to large enterprises. One obstacle in making enterprise networks more energy efficient is the range of devices from multiple vendors deployed on the network. Also, it is difficult to operate with enterprise networks because of their unpredictable growth, frequently changed topology and architecture, regarding the energy consumption.

Current research and developed Network Management Systems (NMS) are not fully automated when executing energy-efficiency policies. The introduction of SDN paradigm brought new opportunities for managing networks through abstraction of lower-level functionality. This means that a centralized SDN controller with only one transaction is able to reconfigure a group of devices. However, the monitoring process of existing NMS to report for changes in the state of the devices is based on point-to-point (p2p) communication with every device on the network. This creates a great deal of traffic and puts additional burden on the network, which raises a sustainable issue in complex networks. Due to improvements in the field of Green IT and smart meters, it is feasible to build a NMS that recons only on power data fetched from the power distribution unit. A pattern for augmenting the values extracted from the power consumption of the network could provide useful information about different network states, for instance detecting changes in the topology. The idea is to use Fault Detection and Isolation (FDI) approach to monitor the network state based on the energy usage. Figure 1 shows that we need two information, a model representing the expected behavior of the devices and the real-time measurements to analyze deviation between the two processes. A deviation corresponds to fault detection in the network, which is in a form of misconfiguration or improper use of the equipment. This means that monitoring the energy consumption could be used not only for Green IT purposes, for raising awareness and reducing the electricity costs, but as well for a classical ICT monitoring system. Also, as the concept of Smart Grid is making use of digital networks to improve the transportation of energy, the work presented in this paper could be explained as the reverse process -how the use of energy could improve the data transport.

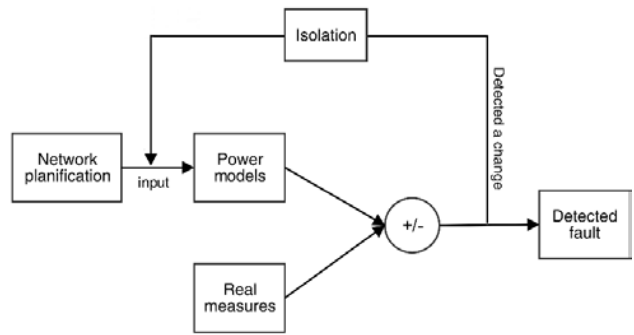

*Figure 1. The general approach of the paper*

The remaining part of this paper is organized as follows: Section 2 presents the related work, while the Section 3 presents the objective of the paper. The system is described in Section 4, which includes the architecture for network management, the design of the experiments and the implementation of the developed application. In Section 5 the obtained results are presented and discussed, while the Section 6 concludes the paper.

**RELATED WORK**

In recent years a new approach for network management emerged under the name of Software Defined Networking (SDN) that allows the network operators to manage the network services through abstraction of high-level functionality. A study by (Kim and Feamster, 2013) proposes an event-driven network control framework based on SDN paradigm and OpenFlow protocol to manage a complex campus network. The focus is on enabling frequent changes to network conditions and states, providing support for network configuration in high-level language, and providing better visibility and troubleshooting. Having a global knowledge of the network state, the developed control framework introduces a centralized approach for network configuration, opposed to distributed management. Meaning that the network operators will not have to configure all the devices individually, but instead let the software make network-wide traffic forwarding decision from a logically single location. Moreover, the network operators provide high-level network policies which are translated by the controller in a centralized manner into a set of forwarding rules, which are used to enforce the policy on the underlying network equipment, by using OpenFlow, as depicted on Figure 2. They offer a set of control domains which can be used by the network operators to define conditions by assigning a suitable packet forwarding actions which corresponds to each condition. Even though the proposed solution reduces the workload of network configuration and management due to the SDN paradigm, the study mainly focuses on the algorithm for translating the policies into a set of reconfigurations of the devices. The system is based on event sources that dynamically collects the current state of the devices, which are inputs to the controller for forming the policies. The event sources monitor the network state and report the changes to the controller, such as bandwidth usage of every end-host device. The monitoring is based on the SNMP and OpenFlow for pulling data with p2p communication which is difficult to manage especially during an expansion of the network, when adding new devices to the network, or during changes to the physical network topology which alters IP address modifications.

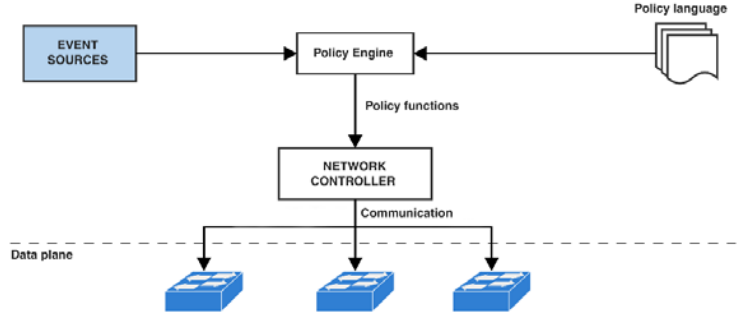

*Figure 2. Classical SDN architecture*

The strategy presented in (Aubrun at al., 2008) for studying the effects of unknown induced delays in network architecture suggests the use of concepts such as Fault Detection and Isolation (FDI) and Fault Tolerance Control (FTC). The study defines a threshold for the expected delay on the basis of the network characteristic and network calculus theory. A faulty situation is then generated and compared to the defined threshold in order to successfully detect which elements are causing the delays and deals with them in a controlled manner. Similarly, (Riekstin et al., 2016) proposes the use of power profiles for each device on the network to determine their expected energy usage under different circumstances. However, the power profiles are considered in special case when the real-time energy measurements are not accessible. They suit as a backup figure to proceed with the FDI's calculations to determine a faulty situation and produce energy efficient policies for the network.

The study by (Drouant et al., 2014) proposes models that gives a global overview of the impact that the network and the ICT equipment is having on the environment, developing the following equation:

$$E = E_m + E_u + E_d$$
$$E_m + \int_{t=0}^{end\ of\ lifecycle} Pu(t)dt + E_d \qquad (1)$$

Where $E_m$ is the energy required for manufacturing and transportation of the equipment, $E_u$ is a factor related to the energy consumed during the usage of the equipment, and the energy required to dismantle the equipment is $E_d$. For the second part of (1), $Pu$ is related to the power consumption by the network architecture during its use phase.

Two strategic approaches exist for modeling the energy consumed during the usage phase of the devices. The first involves a high-level modelling (Foll, 2008) of the whole architecture and the second approach is more precise, meaning that it provides power models for each of the devices part of the network. A high-level model has less interactions with the devices and therefore fails to give accurate estimations how much energy a network architecture consumes at a particular point of time. On the other hand, the use of energy consumption models developed for a particular device would require constant p2p interactions and cause certain amount of additional traffic in the network. However, by giving a more precise figure on the expected behavior of the devices the system would be more responsive to minor changes, faults and anomalies, which is the purpose of the experiments part of this paper. Therefore, the power models for Switch (Reviriego et al., 2012)(Hossain et al., 2015), Router (Ahn and Park, 2014), PC (Agarwal et al., 2009) and Access Point (Demir, Kurt and Karaca, 2014) are added to the Power Model Registry, as shown on the Figure 5, explained in the next sections.

## OBJECTIVE

The goal is to develop sustainable software application suitable for network monitoring systems, which operates only with the retrieved energy consumption data from the ICT devices part of the network infrastructure. The developed application communicates with the Power Grid illustrated on the Figure 3, and by following the FDI approach shown on the Figure 1 is able to detect different states that are occurring on the ICT devices. The network of the power grid includes the Raritan Power Distribution Unit (PDU), which provides the real-time measures for the energy consumption through SNMP periodic pulling. With this, the PDU grid forms the power monitoring process. The detection of the states is achieved by the developed application, which analyzes the retrieved raw power data to distinguish possible faulty situations and anomalies in the usage phase of the ICT devices. The output of the application, or the detected new state suits as an input parameter to the SDN controller. This means that the developed application is envisioned to replace the classical event sources, part of Figure 2, which are using point-to-point communication to report for changes in the state of the devices. To achieve this, the developed software requires two information to be considered: (i) power data from *real-time measurement* procedure and (ii) values for the expected power consumption of each device extracted from the *power models*.

*(i)* The process of *real-time measurement* requires probes and standardized protocols to access the needed metrics which are altogether coupled in a monitoring system. In practice, Raritan Power Distribution Unit (PDU) is the monitored equipment which has a predefined energy Management Information Base (MIB) as a hierarchical structure with properties that is pulled out periodically. The monitoring system itself is then able to collect, analyze and modify the information stored in the MIB through SNMP. It is important to note that the PDU in this case acts as a separate network from the existing network architecture. This means that the monitoring process has a separate channel to the PDU and the packets does not cross the existing network architecture, which releases the burden on the network.

*(ii)* From the other side, as discussed in the section II, the selected *power models* represent the expected behavior of the equipment. Therefore, currently there is an open field to combine the real-time measures, which are translated to the real behavior of a device, with the developed models. A deviation between the models and the monitored power data is used to detect anomalies and to anticipate fault according to a trend analysis.

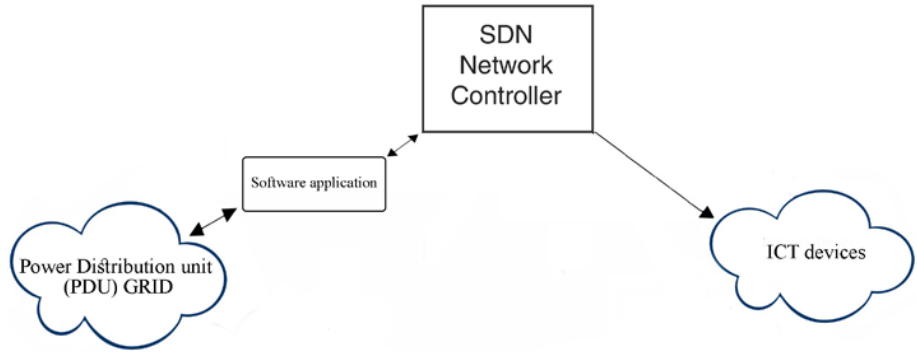

*Figure 3. Objective of the developed system*

The output of the developed system is a continuous detection of changes in the network infrastructure, for instance detecting new device on the network, changes in the Spanning tree protocol, detection in changing the operating state of a switch port, etc. Finally, the objective is the developed monitoring system based on the power grid as shown on the figure 3, which will be a replacement for the existing monitoring system in (Kim and Feamster, 2013) which does a p2p communication with every device part of the network.

**Building a knowledge base**

The developed application that incorporates the monitoring system and the power models has to form thorough foreknowledge in order to successfully anticipate and interpret the readings for the energy consumption. Mostly because the power data is just a raw value and does not contain rich information. As suggested by (Kazandjieva at al., 2013), some types of computing systems within the enterprise can exhibit large variations on the power data even when comparing two instances of the same device model. More precisely, two Dell Optiplex 760 PCs were observed to have over 40% discrepancy in their average power draw. To remedy this, the application has to augment new ways to isolate the problem with the inconsistent energy consumption of the devices, to locate the fault or the anomaly and to classify the problem. One way to deal with this is to predict the fault that may occur during the usage of the equipment. It is important to test the responsiveness of the application by generating different faults and misconfigurations that may occur during a typical workday in an enterprise network. Thus, closely observe the figures for energy consumption and build a knowledge base. A system with such knowledge comprised with smart algorithms is capable of analyzing the deviations and benchmark the faulty situations. From there a pattern could emerge to later detect the problem of aberrant values regarding the energy consumption of the devices during the real-life usage phase of the equipment.

List 1 represents the states which are chosen as most common during a typical enterprise workday, tested on a various heterogeneous combinations of network architectures, prior building the knowledge base. To identify where the faults are applied, a classification based on the objective is depicted on the Figure 4.

**List 1**: states to be detected on the network

*(i)* Testing the operational vs. the sleeping mode on the devices such as switches, routers, access points, LCD monitors, PCs and laptop computers

*(ii)* Turning off/on a switch port

*(iii)* Performing link adaptation on the switch's ports

*(iv)* Evaluating the effect on Energy-Efficient Ethernet (EEE) and Cisco EnergyWise

*(v)* Testing a Spanning Tree Protocol (STP) reevaluation on the network

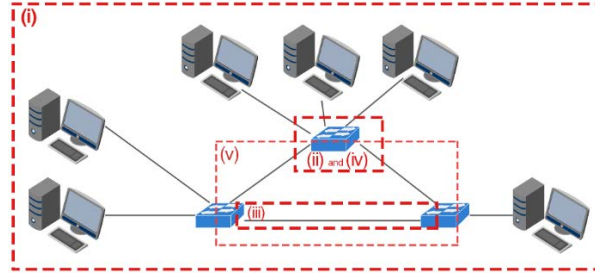

*Figure 4. The classification of the faults*

One possible case study of network architecture is depicted on Figure 5, where the generated faults are represented as the 'Y' input parameter. At disposal for the experiments there were 8 Cisco switches from the series 2960 and 3560, 8 Cisco 1941 Routers, 10 Dell Precision T1700 stations, 2 HP Pavilion laptops, 6 Raritan PX2 PDUs and 2 Raritan PX3 PDUs.

**The logic behind the monitoring system**

The states illustrated on the Figure 6 represent the logic throughout the execution process. Having developed a standalone application able to monitor the energy consumption as a first state allows further analysis process on the retrieved and stored data. The analysis engine as a second state is tightly coupled with the learning model of the system by developed smart algorithms which are able to detect and identify the changes in the power data. Big part of the analysis process plays the inclusion of the power models, discussed in the related work, which are defining the power profiles of the devices and are calculating their expected behavior. The output of the models is then validated with the retrieved real-time measurements for possible deviations. Raritan PDU offers the possibility to monitor its overall consumption as well as the individual power consumption per socket. This means that if a deviation is detected in the comparison process of the overall consumption, the analysis engine will further seek for the device causing the energy disparity. A recent history of energy consumption is retrieved from a database for the detected device and the results are logically passed to the third state, where they are further dismantled towards understanding and interpreting the power data values. The emphasis of the third state, the learning process, is given to the simulation of the faults that are most frequent during the regular usage phase of the ICT equipment. The established knowledge base is used as a baseline on which the learning model is constructed. This means that series of experiments have to be conducted for a range of different network states and faults in the interest of acquiring the desired foreknowledge to benchmark the anticipated states. The benchmarking process is also the key concept for the isolation part as a last state of the logic depicted on the Figure 6.

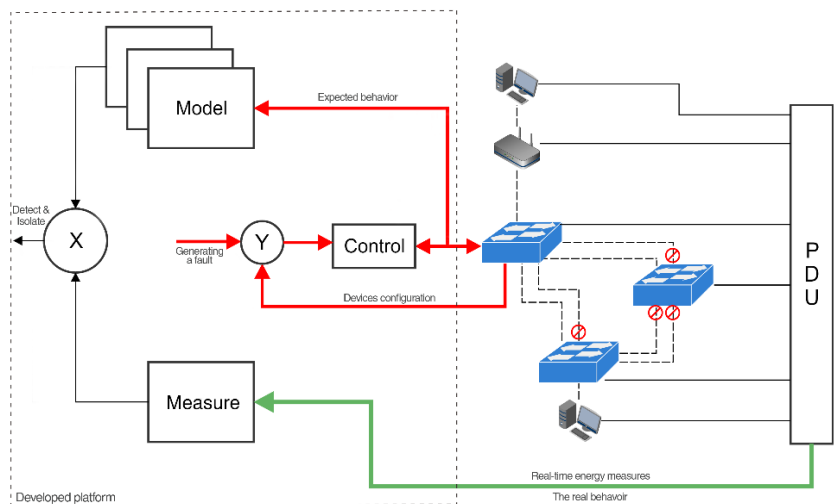

*Figure 5. System behavior*

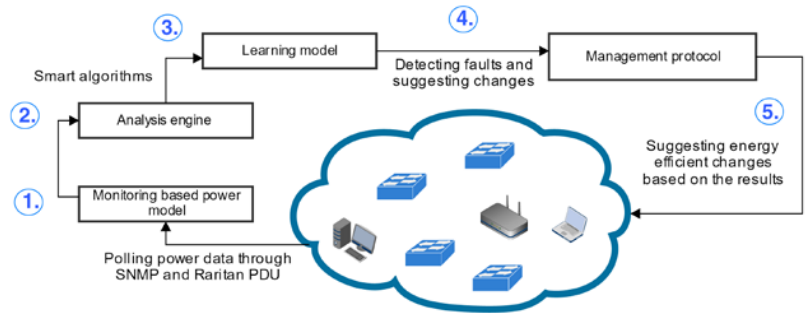

*Figure 6. The experiment's logic*

Ever since the commencement period of the learning process, the system is constantly updating its knowledge base depending on the behavior of the devices, and by following the figures for the energy consumption it is able to discern the aberrant states of the devices. For instance, a possible overloading on the devices, detecting sleep/hibernate states, physical or logical changes in the configuration of the devices, etc. To have seamless operations and accurate output, the learning process also needs to update the static values in the power models based on the detected new state of a particular device. Namely, as seen on the Figure 1, the equations in the power models depend on static parameters from the network planification, which are gathered during one-time measurement. This means that the learning model has to re-compute the input parameters in the power models after a detection of a new network state. For instance, the equation in the power model of a switch has to be updated with new value if a switch port transitioned from offline to online state and vice versa. The outcome of the learning process regarding the benchmarking part concludes the fourth state of the logic behind the experiment, the engine for power management. An accurate benchmarking provides the possibility to identify the changes that occurred on the network infrastructure and suggests a recuperation, or a reconfiguration of the network.

# SYSTEM DESCRIPTION

This section provides more technical explanation of the developed application. The architecture of the system has been previously described in (Minovski, Rondeau and Georges, 2016).

**Implementation**

The application, which includes the power models and the monitoring module is implemented in Java language using the SNMP library and MySql database. The energy monitoring task is developed in a multi-threaded fashion to periodically pull the energy values from the Raritan PDU, by importing Java libraries for SNMP to access the functions for requesting the right Raritan's MiB OIDs. There is an ability to optimize the frequency of the probes send to the Raritan PDU throughout a developed GUI, thus the accuracy of the measures is practically defined by the operator. During the test phase of the application, probes were requested every second and the responses are stored locally with the interest to enable further analysis on the data. The response from each probe is a group of values in the following format: *$I_{(A)}$ – Electric current significant with $10^{-3}$ amps (A) base; $V_{(V)}$ – Voltage; PF – Power factor*. To get the current energy consumption (W) the following equation has to be performed:

$$W = I_{(A)} \times V_{(V)} \times PF \qquad (2)$$

Moreover, the probe response does not only contain data for the total energy consumed by the whole PDU, but also it reports the consumption of each socket individually. This means that the application with additional analysis can successfully build its own schemas and detect what kind of device is connected to a certain socket. Giving this ability, the process flow can successfully discover which device specifically is not consuming the energy as it is expected. Algorithm 1 describes the process flow of the application.

---

**Algorithm 1**: the process flow

input: PM, the power models; PR, the probe response
output: $F_{fault}$, detected fault, $I_{isolation}$, isolating the fault

*R is the Raritan PDU*
*E denotes the attached equipment*
*KB represents the knowledge base in a form of database*

**foreach** X in PR **do**
    $W_{total}$ ← $X(I_{(A)}) * X(V) * X(PF)$
    Store($W_{total}$)
    **if** $W_{total}$ > $PM_{total}$ **then**
        **foreach** E in R **do**
            $W_{individual}$ ← $E(I_{(A)}) * E(V) * E(PF)$
            Store($W_{individual}$)
        **if** $W_{individual}$ > $PM_{individual}$ **then**
            $F_{fault}$ ← Detection(E, KB, History($W_{individual}$))
            $I_{isolation}$ ← Isolation($F_{fault}$)
            Store($F_{fault}$, $I_{isolation}$)
**return** $F_{fault}$, $I_{isolation}$

The support of power models (PM) in the application creates the premises to set the expected threshold regarding the energy spent by each detected device. The discussed power models are represented with a static parameters gathered by one-time measurement at the initialization phase of the application or when a new device is plugged to the PDU socket.

Having a real-time availability probes enables the application to perform parallel comparison on the retrieved data for individual sockets, and as a whole, with the data provided by the power models. In case of aberrant values, the application tries to locate the device with the unexpected behavior, load the latest stored power data history of the device and perform analysis for fault detection. The FDI concept is based on calculations directed by the predefined fault foreknowledge, achieved by the experiments for generating the faults in order to benchmark the consumption that a particular occurring state is causing. The knowledge base (KB) is also integrated in the algorithm in a form of a database.

The Detection() function is the core of the algorithm due to the fact that it couples the analysis and the learning process, described in the previous section. It begins by locating the device that is observed to have disparity in the energy consumption. Also, a recent history of the spotted device is passed to the function, which includes two separate components. First is the recent energy consumption history of the device and secondly is the recent history of the noted running states and configuration of the device. This process maps the previous running configuration and understands the current states of the devices that might be causing the energy disparity. Moreover, the knowledge base is as well loaded which can interpret the difference in the expected and the real energy measures into a new network state. The function combines the above mentioned parameters and the calculations result with detection of the changed state. For instance, a switch is detected to have powered off a port, changed the link speed, or perhaps reevaluated the STP.

The Isolation() function as a input parameter receives the detected new state and performs further analyses to examine the implications, especially on the energy consumption. The analysis determines the actions that have to be taken upon detection of new state. For instance, if a switch port is detected to transition from online to offline mode, this function re-computes the input parameters of the static power models to make them having accurate calculations of the expected energy consumption of the switch. This means that the examination of the implications result with discovery whether the detected new state is a misconfiguration of the network, a faulty situation or just changes in the network state that have to be noted. As an output, the Isolation() function updates the database of the spotted device recent history and the knowledge base, and reports to the network operator for the changes.

The first version of the GUI is show on the Figure 7. The network operator is able to see live chart of the energy consumption of the whole network, as well as select specific PDU and monitor the consumption. Based on the detected states, the network operator could map the running configuration of the whole network, as well as separately for each PDU. When a specific PDU is selected, the operator could monitor the individual sockets, see the detected device and read the changes in the state. There is a possibility to download a report with a complete history of the changes on the network. The obtained results are discussed in the next section.

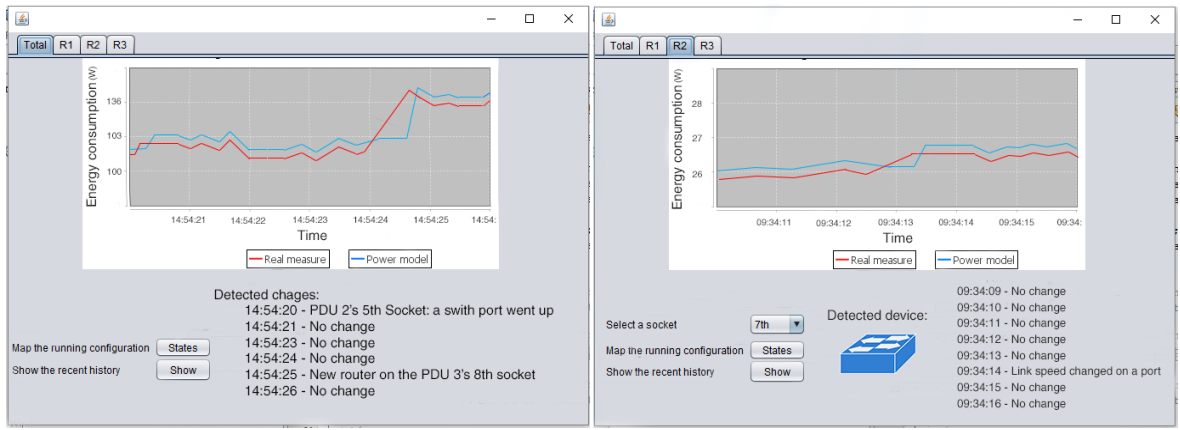

*Figure 7. Developed GUI*

## RESULTS AND DISCUSSION

This section refers to the List 1, presenting the obtained results during the test phase of the developed application. Each of the experiments were tested on different network architectures, monitored for 10 minutes continuously, as well as under different amount of load generated to flow through the network.
*(i)* The possibility to automatically detect the working hours of an enterprise allows the application dynamically to report on the equipment that was left in an operational state. For instance, if a wireless access point is observed to consume constant level of energy during non-working hours, the application will indicate that the device is powered, but does not transmit any data, meaning that it should transition into a sleep mode. The cost of powering the hardware components of each device dominates the overall power profile and therefore the opportunity for energy savings are the highest. The difference in the energy consumption between the operational and the sleeping mode is illustrated on the Figure 8. Typically, there is a burst of energy consumption when a device goes to operational state, either from sleep/hibernate or offline mode. Those values are also used to benchmark the current state of the device.

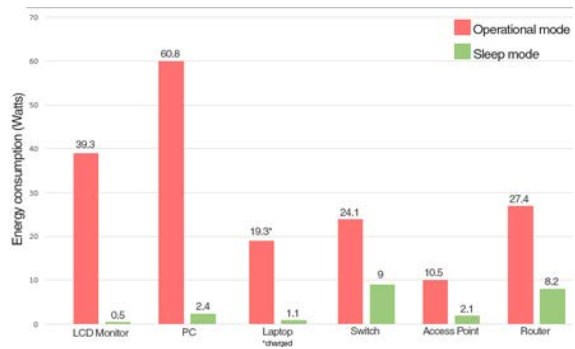

*Figure 8. Obtained results – experiment (i)*

*(ii)* Managing the active ports on a switch is one important and exquisite objective because it is the only way of providing network accessibility. Therefore, to have an overview on the status of the ports at one point of the time by observing the switch's values for the energy consumption is a delicate task. Mainly because of the introduction of EEE as a set of enhancements able to dynamically put a single link temporarily to sleep when not in use. The difference in the power consumption after a transition from active to inactive switch port is observed to be from 0.3 to 0.4 W, as illustrated on the Figure 9. This result allows the developed application to benchmark the state and have detailed matrix of the port statuses of each switch. Having this feature allows issuing reports for possible unnecessarily active ports, but also is part of the analysis for a possible misconfiguration of the network.

*(iii)* The utilization of the Ethernet links is, on average, extremely low (Gunaratne et al., 2008). This suggests that there is an ample opportunity for energy savings by operating Ethernet links at a low data rate for most of the time with no perceivable performance impact to the user. To keep track of the each switch's port assigned speed, by following the energy consumption, means an opportunity for adjustments according to the needs. But it also helps the application to correlate the assigned link speed with the congestion, or the load on the switch, which is as well affecting the overall power consumption. During the observations, there is no notable change in the consumption when comparing 10Mb/s to 100Mb/s speed link, but the difference when making a transition from 1Gb/s to 100Mb/s is from 0.2 to 0.4 W per port. Thus, the developed application by benchmarking this value is forming a schema of each port link speed that is assigned. The outcome is a suggestion when to reduce the speed of a link when a low utilization is perceived.

*(iv)* Energy-Efficient Ethernet is not yet a feature supported by many network devices currently on the market, however it is important to observe its impact on the energy consumption. During an idle period of a switch with no traffic flowing through the ports, EEE will put all of the active ports in the Low Power Idle (LPI) state, and therefore achieve the same result as the second *(ii)* experiment of 0.3 to 0.4 W savings per port. Hence, the developed application has to be aware of the possibilities of EEE to dynamically, for a temporary time, put a port to idle mode.

*(v)* Spanning Tree Protocol (STP) is commonly used Layer 2 protocol that runs on switches and bridges, with purpose to ensure that there are no loops created on a redundant paths in the network. By default STP dynamically manages the elimination of the loops with the process of electing a tree based on priorities with a particular switch as a root. This means that there is a possibility of a network reconfiguration without the assistance of the operator, which could lead to additional power demands if it is not well managed. STP, enabled between the switches on the Figure 6, does not save any energy with its ability to block certain port in order to discard the loops in the network. Namely, a port in a blocking state is consuming the same amount of 0.4 W as the other active ports. The only noticeable difference is the spike of energy consumption when STP is trying to re-evaluate the tree. Namely, during a physical or a logical generated error, STP will elect a new root and construct the new tree, thus will add up to 1 W for few seconds to the overall consumption of the switch, depending on the architecture. This burst of energy, also shown on the Figure 9, is used to benchmark a change in the STP, with the interest of reporting it for possible misconfiguration of the network.

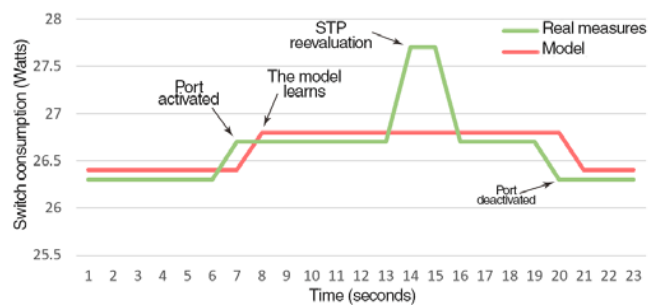

*Figure 9. Obtained results – experiments (ii) and (v)*

# CONCLUSION AND FUTURE WORK

This paper presents the achieved results from evaluating the proposed novel extension for energy-efficient network management application emulated in a heterogeneous network environment. The results shows that a reliable application can be designed to monitoring the changes in the state of the devices with minor impact on the network regarding the traffic. The proposed monitoring process is designated as a replacement for the existing p2p pulling process, part of the SDN typical systems. The SDN controller, and thus the network operator, has the ability to receive reports for the changed state of the network in an energy-efficient way by following the energy consumption of the network infrastructure. The approach of monitoring the power consumption directly from the PDU, instead of p2p communication with every device, brings own complexity as the network grows. For instance, in a scenario of multiple faults happening simultaneously, the FDI concept presented in this paper is able to cope with the occurring situation. This is achieved by the possibility of following each PDU's socket individually from a centralized application. The support for real-time availability calculation enables the operator to take service-centric decisions, rather than network centric as proposed currently in the literature.

As future work, we envision a deployment of the proposed application as part of a SDN controller. Moreover, the application should be scalable in terms of network size and adapt to any SDN software controller developed on different programming language, with diverse libraries. The anticipation is also to include other common network anomalies and to follow their impact on the energy consumption, for instance the packet loss rate and highly congested networks, which could possibly trigger a use of other techniques in order to gather all the necessary information for the analysis process, rather than just monitoring the energy values.

# ACKNOWLEDGEMENT

This work is part of the Erasmus Mundus Master programme in Pervasive Computing and Communication for Sustainable Development (PERCCOM) (Klimova et al., 2013) of the European Union (www.perccom.eu). The authors thank all the partner institutions, sponsors and researchers of the PERCCOM programme.